# High-pressure synthesis of bilayer nickelate $Sr_3Ni_2O_5Cl_2$ with tetragonal crystal structure

**Kazuki Yamane[a][b]\*, Yoshitaka Matsushita[c], Shintaro Adachi[d], Ryo Matsumoto[a], Kensei Terashima[a], Takanobu Hiroto[c], Hiroya Sakurai[a] and Yoshihiko Takano[a][b]**

[a]International Center for Materials Nanoarchitectonics (MANA), National Institute for Materials Science, 1-2-1 sengen, Tsukuba, Ibaraki, 305-0047, Japan

[b]Graduate School of Pure and Applied Sciences, University of Tsukuba, 1-1-1 Tennodai, Tsukuba, Ibaraki, 305-8577, Japan

[c]Research Network and Facility Services Division (RNFS), National institute for Materials Science, 1-2-1 Sengen, Tsukuba, Ibaraki, 305-0047, Japan

[d]Nagamori Institute of Actuators, Kyoto University of Advanced Science, 18 Gotanda, Yamanouchi, Ukyo, Kyoto, 615-8577, Japan

Correspondence email: yamane.kazuki@nims.go.jp

**Synopsis** A theoretical candidate for a new Ni oxide superconductor, $Sr_3Ni_2O_5Cl_2$, was synthesized for the first time under high-pressure conditions (10 GPa, 1400°C). While it adopts the tetragonal Ruddlesden—Popper phase structure predicted by theoretical calculations, resistance measurements revealed no superconductivity down to 2 K under pressures up to 24 GPa.

**Abstract** A novel oxychloride, $Sr_3Ni_2O_5Cl_2$, was synthesized for the first time under high pressure of 10 GPa at 1400°C, motivated by a theoretical prediction of its potential superconductivity under ambient pressure. Small single crystals were used to determine the crystal structure and measure the temperature dependence of electrical resistance. The crystal is isostructural with the recently discovered superconductor, $La_3Ni_2O_7$, in line with the theoretical expectation.

# 1. Introduction

Recently, La$_3$Ni$_2$O$_7$ has been reported to exhibit superconductivity under pressures exceeding 15 GPa (Sun *et al*., 2023). Remarkably, the superconducting transition temperature reaches as high as 80 K, comparable to that of high-$T_c$ cuprates. The compound adopts a Ruddlesden−Popper phase structure, and the superconductivity occurs in the double layers of NiO$_2$ square lattices like the superconductivity in CuO$_2$ square lattices in the cuprates. These intriguing characteristics have motivated the search for other Ni oxide superconductors (Sakakibara *et al*., 2024*b*; Nagata *et al*., 2024; Ueki *et al*., 2024), with the hope that one might exhibit superconductivity at ambient pressure. This would achieve a significant breakthrough in the field.

The superconductivity of La$_3$Ni$_2$O$_7$ had been theoretically predicted even before its experimental discovery, as the intercoupling between two square lattices in the double layer is thought to be advantageous for superconductivity (Nakata *et al*., 2017; Sakakibara *et al*., 2024*a*; Kaneko *et al*., 2024). Thus, the key to the occurrence of superconductivity is widely believed to lie in the Ni—O$_{ap}$—Ni bridging angle (Sun *et al*., 2023), where O$_{ap}$ represents the oxygen ion between two Ni ions along the *c*-axis. In fact, the superconductivity appears above the pressure at which the orthorhombic *Amam* structure transforms into the tetragonal *I*4/*mmm* structure (Wang *et al*., 2024*a,b*). This transformation causes the bridging angle to change from 168° to 180°, presumably enhancing the intercoupling between the two square lattices.

Based on the same theoretical approach used to predict the superconductivity of La$_3$Ni$_2$O$_7$, the previously unreported compound Sr$_3$Ni$_2$O$_5$Cl$_2$ has recently been proposed as a promising candidate for a superconductor under ambient pressure (Ochi *et al*., 2024). It is expected to exhibit tetragonal *I*4/*mmm* symmetry even under ambient pressure, if it exists. Therefore, we decided to synthesize this new Ni oxychloride and successfully obtained a single crystal of it. In this report, we present the method of the synthesis, crystal structure, and electrical resistance measurements.

## 2. Experimental

Single crystals of $Sr_3Ni_2O_5Cl_2$ were synthesized via high-pressure techniques using a stoichiometric mixture of $SrO_2$, SrO, $SrCl_2$, and NiO in a Kawai-type multi-anvil press (Kawai *et al*., 1970). $SrO_2$ was prepared by precipitation from a reaction between $Sr(NO_3)_2$ and $H_2O_2$. Specifically, $Sr(NO_3)_2$ was dissolved in deionized water that had been pre-degassed by bubbling Ar gas to remove dissolved $CO_2$. A 30% aqueous $H_2O_2$ solution was then added, followed by aqueous $NH_3$ to induce precipitation. The precipitate was filtered and dried at 150°C. All steps were carried out in a glove bag filled with Ar gas to minimize $CO_2$ exposure. SrO was obtained by thermal decomposition of $SrCO_3$ at 1200°C under flowing Ar gas, while $SrCl_2$ was prepared by dehydrating $SrCl_2 \cdot 6H_2O$ at 300°C under vacuum. The stoichiometric mixture was sealed in a Pt capsule inside an Ar-filled glove box, heated to 1400°C at 10 GPa for two hours in the press, and then the temperature was reduced to room temperature by lowering it in 100 °C increments, with a 20-minute hold at each step, before releasing the pressure. The final product included single crystals with typical dimensions of 50 μm × 50 μm × 30 μm.

Single crystals were examined using a scanning electron microscope (SEM, JSM-6010LA, JEOL) equipped with energy-dispersive X-ray spectroscopy (EDX) for elemental analysis. Single-crystal X-ray diffraction (SCXRD) data were collected at 300 K using a RIGAKU Synergy Custom DW single crystal diffractometer with VariMax confocal optics for Mo K$\alpha$ radiation ($\lambda = 0.71073$ Å) and a HyPix2000 detector. A crystal with dimensions of approximately $30 \times 27 \times 19$ μm$^3$ was isolated under paraffine oil, then immediately mounted in a dry $N_2$ gas stream to prevent degradation, as the compound is highly air-sensitive. Cell refinement and data reduction were carried out by using the program CrysAlis Pro. Preliminary structures were solved using SHELXT (Sheldrick *et al*., 2015) and refined by full-matrix least squares on $F^2$ using the SHELXL-2018/3 (Sheldrick *et al*., 2015) in Olex2 program package (Dolomanov *et al*., 2009). Electrical resistance measurements under pressures were performed using a diamond anvil cell (DAC) equipped with boron-doped diamond electrodes directly fabricated onto the diamond surface (Matsumoto *et al*., 2016; Sakakibara *et al*., 2024*b*). Resistance data were obtained with a Physical Property Measurement System (PPMS, Quantum Design).

## 3. Results and discussion

Figure 1 shows the SEM image of a representative crystal. The plate-like morphology is consistent with the layered structure typical of the Ruddlesden—Popper phase, indicating that the crystal readily cleaves along these planes. Notably, the depicted crystal was cleaved using Scotch tape, highlighting the material's inherent cleavage planes. EDX analysis determined the composition to be $Sr_{3.1}Ni_{2.1}O_xCl_{1.8}$, in close agreement with the nominal stoichiometry of the starting materials.

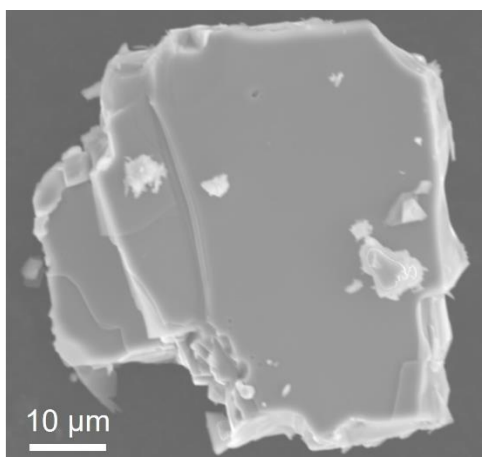

**Figure 1** SEM image of a single crystal of $Sr_3Ni_2O_5Cl_2$.

The parameters obtained from the SCXRD structure refinements are summarized in Tables 1 and S1. The final cycle of refinement, performed on $F^2$ with 18 variables, 448 averaged reflections [$F^2 > 2\sigma (F^2)$], and 500 reflections [$F$], converged to residual values of $R[F^2>2\sigma(F^2)] = 0.0285$ and $wR(F^2) = 0.0757$. No significant deviations from full occupancy were observed for any of the atoms. Consequently, the crystal structure was determined to be isostructural with $La_3Ni_2O_7$ and $Sr_3M_2O_5Cl_2$ (M = Sc, Fe, and Co) (Wang *et al*., 2024; Su *et al*., 2018; Leib et al., 1984; Mcglothlin et al., 2000). Specifically, the compound crystallizes in the Ruddlesden—Popper phase structure, adopting the tetragonal *I*4/*mmm* symmetry with lattice parameters $a = 3.83990(10)$ Å, $c = 24.2936(12)$ Å, as shown in Fig. 2 (*a*).

In the structure, there are two oxygen sites, 2*a* and 8*g*, referred to as $O_{ap}$ and $O_{eq}$, respectively, in this paper. $O_{ap}$ bridges two Ni atoms along the *c*-axis, while $O_{eq}$ is located between two Ni atoms in the *ab*-plane. The bond lengths for Ni—$O_{ap}$, Ni—$O_{eq}$, Ni—Cl are 1.8718(5) Å,

1.9389(4) Å, and 3.1023(14) Å, respectively. Based on these bond lengths, we have estimated the valence state of atoms using the bond-valence-sum (BVS) method (Brown *et al*., 1985; Brese *et al*., 1991). The BVS parameters of $Ni^{2+}$—$Cl^-$ of 2.02 was used (Brese *et al*., 1991), following a previous literature procedure for the related compound $Sr_2NiO_3Cl$ (Tsujimoto *et al*., 2013).

The Ni valence was estimated to be +3.17. Similarly, the valences of Sr1 and Sr2 were estimated from the coordination environments, shown in Fig. 2 (*d*), to be +1.83 and +2.57, respectively. These values are in reasonable agreement with the formal valence expected from the composition, considering that the structure is stabilized only under high-pressure conditions (Ochi et al., 2024). Another high-pressure phase of Ni oxide, $Sr_2NiO_3Cl$, (Tsujimoto *et al*., 2013) also exhibits a larger Sr valence (+2.51) at the Sr2 site, which is surrounded by both O and Cl ions as shown in Fig. 2 (*f*). The asymmetry in coordination may account for the higher BVS valences. In contrast, the Ni ions are displaced toward the $O_{ap}$ ions from the center of the $O_{eq}$ coordination plane, reducing the Ni—$O_{eq}$—Ni bridging angle from 180° to 164.0° in $Sr_3Ni_2O_5Cl_2$. This value closely matches the theoretical prediction of 162°, further reinforcing the validity of the theoretical calculation.

Interestingly, the Ni—$O_{ap}$ bond length is significantly shorter than the Ni—$O_{eq}$ bond length. Although this trend is also observed in $Sr_2NiO_3Cl$ as shown in Fig. 2 (*e*), it is more pronounced in $Sr_3Ni_2O_5Cl_2$. In fact, the ratio of Ni—$O_{ap}$ to Ni—$O_{eq}$ is 0.965 in $Sr_3Ni_2O_5Cl_2$ whereas that of $Sr_2NiO_3Cl$ is 0.975, suggesting stronger covalency between Ni—$O_{ap}$ in $Sr_3Ni_2O_5Cl_2$ than in $Sr_2NiO_3Cl$. The larger displacement of the transition metal cations toward $O_{ap}$ in the double-layered structure, compared to the single-layered structure, may be a common characteristic feature of strontium transition-metal oxychlorides (Hector *et al*., 2001; Leib *et al*., 1984; Loureiro *et al*., 2000). Thus, the highly enhanced vertical interlayer hopping in $Sr_3Ni_2O_5Cl_2$ predicted by the theory is presumably attributed to this feature of the strontium transition-metal oxychlorides, being related to the high covalence between Ni and $O_{ap}$.

**Table 1** Wyckoff positions (WP), occupancy, fractional atomic coordinates, isotropic and anisotropic atomic displacement parameters (Å$^2$) of Sr$_3$Ni$_2$O$_5$Cl$_2$.

| Atom | WP | Occ. | x | y | z | $U_{iso}$ | $U_{11}$ | $U_{22}$ | $U_{33}$ |
|---|---|---|---|---|---|---|---|---|---|
| Sr1 | 2a | 1 | 0 | 0 | 0.5 | 0.01719(12) | 0.01381(13) | = $U_{11}$ | 0.0239(3) |
| Sr2 | 4e | 1 | 0 | 0 | 0.34453(2) | 0.01499(10) | 0.01223(10) | = $U_{11}$ | 0.02051(19) |
| Ni | 4e | 1 | 0 | 0 | 0.07705(2) | 0.01226(11) | 0.01015(13) | = $U_{11}$ | 0.01650(2) |
| O$_{ap}$ | 2a | 1 | 0 | 0 | 0 | 0.0193(8) | 0.02260(12) | = $U_{11}$ | 0.01280(17) |
| O$_{eq}$ | 8g | 1 | 0 | 0.5 | 0.08818(10) | 0.0171(4) | 0.0156(8) | 0.0101(7) | 0.0256(10) |
| Cl | 4e | 1 | 0 | 0 | 0.24750(5) | 0.0208(2) | 0.0181(3) | = $U_{11}$ | 0.0263(5) |

*) $U_{12} = U_{13} = U_{23} = 0$

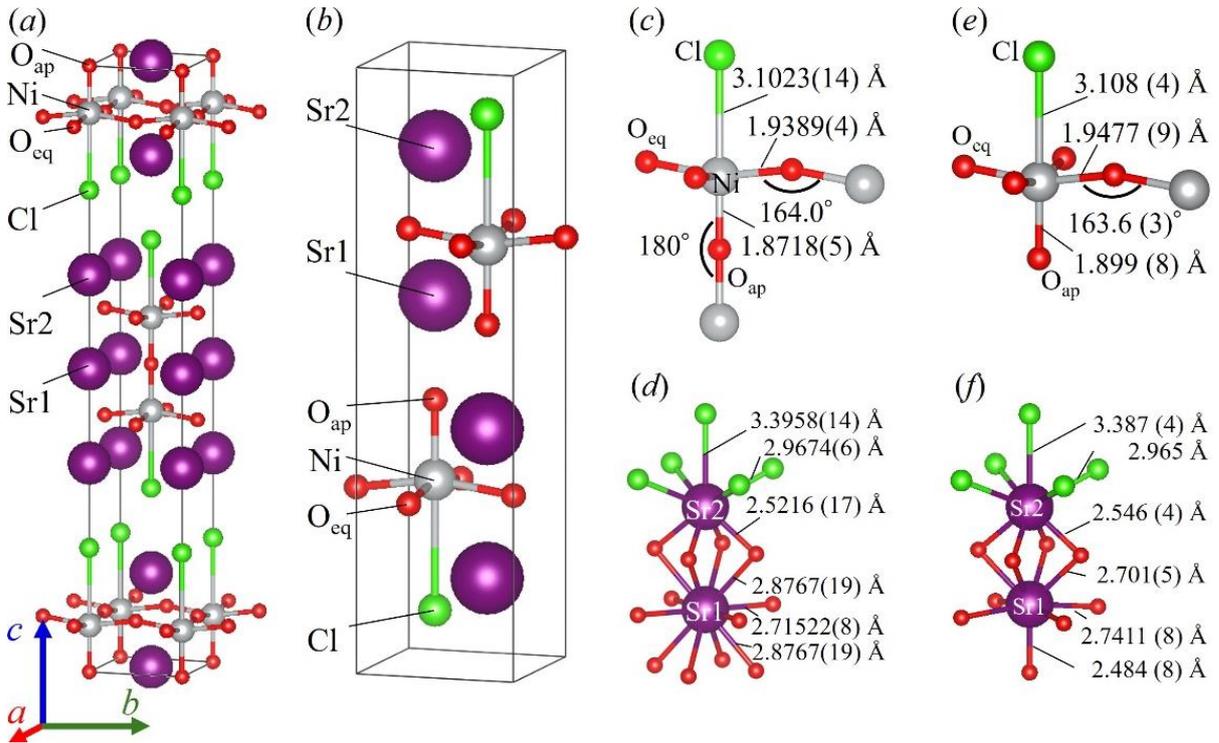

**Figure 2** Crystal structures of Sr$_3$Ni$_2$O$_5$Cl$_2$ (*a*) and Sr$_2$NiO$_3$Cl (*b*) (Tsujimoto et al., 2013), and local coordination environments around a Ni site (*c*) and Sr sites (*d*) in the former compound, and those in the latter compound (*e*, *f*). The boxes in the panels a and b represent unit cells.

The temperature dependence of electrical resistance under various pressures are shown in Fig. 3. Although the resistance increases with decreasing temperature at 0.2 GPa, which is almost the ambient pressure, the compound remains highly conductive, consistent with the metallic nature predicted by the theory. However, no superconductivity is observed down to 2 K. The increase in resistance with decreasing temperature is gradually supressed by applying pressure up to 24 GPa, However, the superconductivity is not emerged. The origin of the absence of the superconductivity is under investigation.

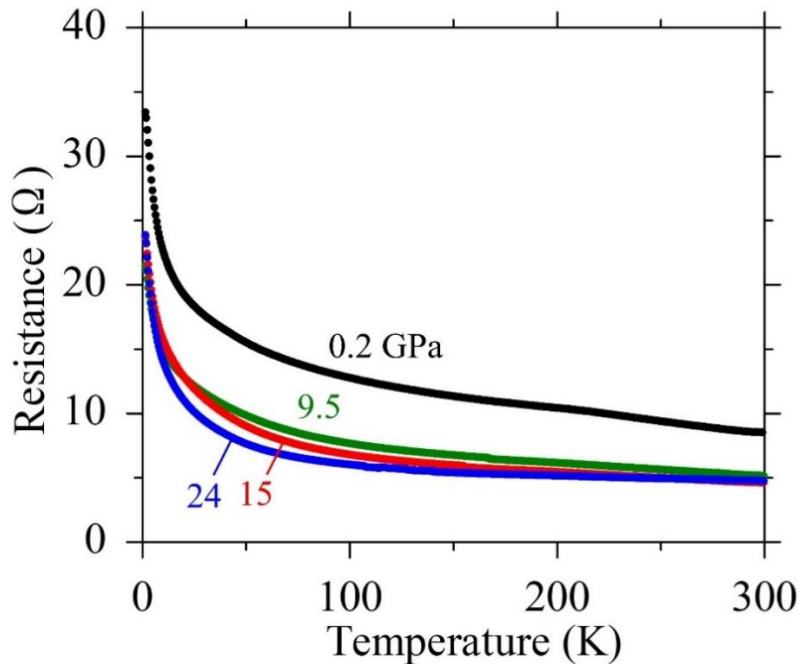

**Figure 3** Temperature dependence of electrical resistance of $Sr_3Ni_2O_5Cl_2$ under various pressures.

**Acknowledgements**   We thank Profs. Masayuki Ochi (Osaka Univ.), Hirofumi Sakakibara (Tottori Univ.), Hidetomo Usui (Shimane Univ.), and Kazuhiko Kuroki (Osaka Univ.) for fruitful discussions. This work is supported by World Premier International Research Center Initiative (WPI), MEXT, Japan, and JSPS KAKENHI Grant Nos. JP20H05644 and JP24K01333. VESTA was used to generate Figs. 3 (Momma & Izumi, 2011).

**Conflicts of interest**   The authors declare no competing interests.

**Data availability**   The authors confirm that the data supporting the findings of this study are available within the article and its supplementary materials.

**References**

Brese, N. E., & O'Keeffe, M. (1991). *Acta Crystallogr*. B**47**, 192-197.

Brown, I. D., & Altermatt, D. (1985). *Acta Crystallogr*. B**41**, 244-247.

CrysAlis Pro ver. 1.171.43.125a, Rigaku Oxford Diffraction, Tokyo, Japan (2024).

Dolomanov, O. V., Bourhis, L. J., Gildea, R. J., Howard, J. A. K., & Puschmann, H. (2009). *J. Appl. Crystallogr.* **42**, 339-341.

Hector, A. L., Hutchings, J. A., Needs, R. L., Thomas, M. F., & Weller, M. T. (2001). *J. Mater. Chem*. **11**, 527-532.

Kaneko, T. Sakakibara, H. Ochi, M. & Kuroki, K. (2024). *Phys. Rev. B* **109**, 045154.

Kawai, N., & Endo, S. (1970). *Rev. Sci. Instrum.* **41**, 1178-1181.

Leib, W., & Müller-Buschbaum, Hk. Z. (1984). *Anorg. Allg. Chem.* **518**, 115-119.

Loureiro, S. M., Felser, C., Huang, Q., & Cava, R. J. (2000). *Chem. Mater.* **12**, 3181-3185.

Matsumoto, R., Sasama, Y., Fujioka, M., Irifune, T., Tanaka, M., Yamaguchi, T., Takeya, H., & Takano, Y. (2016). *Rev. Sci. Instrum.* **87**, 076103.

Mcglothlin, N., Ho, D., & Cava, R., J. (2000). *Mater. Res. Bull*. **35**, 1035-1043.

Momma, K., & Izumi, F. (2011). *J. Appl. Crystallogr*. **44**, 1272–1276.

Nagata, H., Sakurai, H., Ueki, Y., Yamane, K., Matsumoto, R., Terashima, K., Hirose, K., Ohta, H., Kato, M., & Takano, Y. (2024). *J. Phys. Soc*. Japan **93**, 095003.

Nakata, M., Ogura, D., Usui, H., & Kuroki, K. (2017). *Phys. Rev. B* **95**, 214509.

Ochi, M. Sakakibara, H. Usui, H. & Kuroki, K. (2024). *arXiv* 2409.06935.

Sakakibara, H., Kitamine, N., Ochi, M., & Kuroki, K. (2024*a*). *Phys. Rev. Lett*. **132**, 106002.

Sakakibara, H., Ochi, M., Nagata, H., Ueki, Y., Sakurai, H., Masumoto, R., Terashima, K., Hirose, K., & Ohta, H. (2024*b*). *Phys. Rev. B* **109**, 144511.

Sheldrick G., M. (2015). *Acta Crystallogr*. **A71**, 3–8.

Sheldrick G., M. (2015). *Acta Crystallogr*. **C71**, 3–8.

Sun, H., Huo, M., Hu, X., Li, J., Liu, Z., Han, Y., Tang, L., Mao, Z., Yang, P., Wang, B., Cheng, J., Yao, D. X., Zhang, G. M., & Wang, M. (2023). *Nature* **621**, 493-498.

Su, Y., Tsujimoto, Y., Fujii, K., Tatsuta, M., Oka, K., Yashima, M., Ogino, H., & Yamaura, K. (2018). *Inorg. Chem.* **57**, 5615-5623.

Tsujimoto, Y., Yamaura, K., & Uchikoshi, T. (2013). *Inorg. Chem.* **52**, 10211-10216.

Ueki, Y., Sakurai, H., Nagata, H., Yamane, K., Matsumoto, R., Terashima, K., Hirose, K., Ohta, H., Kato, M., Takano, Y. (2024). *arXiv* 2408.04970.

Wang, L., Li, Y., Xie, S. Y., Liu, F., Sun, H., Huang, C., Gao, Y., Nakagawa, T., Fu, B., Dong, B., Cao, Z., Yu, R., Kawaguchi, S. I., Kadobayashi, H., Wang, M., Jin, C., Mao, H. K., & Liu, H. (2024*a*). *J. Am. Chem. Soc.* **146**, 7506-7514.

Wang, N., Wang, G., Shen, X., Hou, J., Luo, J., Ma, X., Yang, H., Shi, L., Dou, J., Feng, J., Yang, J., Shi, Y., Ren, Z., Ma, H., Yang, P., Liu, Y., Zhang, H., Dong, X., Wang, Y., Jiang, K., Hu, J., Nagasaki, S., Kitagawa, K., & Cheng, J. (2024*b*). *Nature* **634**, 579-584.

**Supporting information**

**Table S1** Structure refinement of $Sr_3Ni_2O_5Cl_2$

---

*Crystal data*

| | |
|---|---|
| $Sr_3Ni_2O_5Cl_2$ | $D_x$ = 4.925 |
| $Mr$ = 531.18 | Mo Kα radiation, $\lambda$ = 0.71073 |
| Tetragonal, $I4/mmm$ (#139) | Cell parameters from 4484 reflections |
| $a$ = 3.8399 (1) | $\theta$ = 3.327 – 44.742° |
| $c$ = 24.2936 (12) | $\mu$ = 28.061 mm$^{-1}$ |
| $V$ = 358.21 (3) | $T$ = 300.8(9) K |
| $Z$ = 2 | block, metallic black |
| $F$ (000) = 488 | ~30 × 27 × 19 μm |

*Data collection*

| | |
|---|---|
| ROD, Synergy Custom DW system, HyPix Diffractometer | 500 Independent reflections |
| | 448 reflections with $[F^2 > 2\sigma(F^2)]$ |
| Detector resolution: 10.0000 pixels mm$^{-1}$ | $R_{int}$ = 0.0649 |
| $\omega$ scans | $\theta_{max}$ = 44.860°, $\theta_{min}$ = 3.354° |
| Absorption correction: multi-scan | $h$ = −7→7 |
| (CrysAlisPro 1.171.43.125a (Rigaku Oxford | $k$ = −7→7 |
| Diffraction, 2024)) | $l$ = −48→48 |
| $T_{min}$ = 0.34623, $T_{max}$ = 0.37985 | |
| 8872 measured reflections | |

*Refinement*

| | |
|---|---|
| Refinement of $F^2$ | $w = 1/[\sigma^2(F_o^2)+(0.0395P)^2+0.6699P]$ |
| Least-squares matrix: full | where $P = (F_o^2+2F_c^2)/3$ |
| $R[F^2>2\sigma(F^2)]$ = 0.0285 | $(\Delta/\sigma)_{max}$ < 0.002 |
| $wR(F^2)$ = 0.0757 | $\Delta\rho_{max}$ = 3.442 e Å$^{-3}$ |
| $S$ = 1.123 | $\Delta\rho_{min}$ = −1.003 e Å$^{-3}$ |
| 500 reflections | |
| 18 parameters | |
| 0 restraints | |
| Primary atom site location: dual | |